\documentclass[pre,aps,preprint,showpacs]{revtex4} 
\usepackage{revsymb,amsmath}
\usepackage{graphicx}
\begin{document}
\author{Thomas Michael, Steffen Trimper}
\affiliation{Fachbereich Physik,
Martin-Luther-Universit\"at,D-06099 Halle, Germany}
\email{trimper@physik.uni-halle.de}
\author{Michael Schulz}
\affiliation{Abteilung Theoretische Physik, Universit\"at Ulm, D-89069 Ulm, Germany}
\email{michael.schulz@physik.uni-ulm.de}
\title{On the Glauber model in a quantum representation }
\date{\today }

\begin{abstract}
The Glauber model is reconsidered based on a quantum formulation of the Master equation. 
Unlike the conventional approach the temperature and the Ising energy are included from the 
beginning by introducing a Heisenberg-like picture of the second quantized operators. This 
method enables us to get an exact expression for the transition rate of a single flip-process 
$w_i(\sigma _i)$ which is in accordance with the principle of detailed balance. The transition 
rate differs significantly from the conventional one due to Glauber in the low temperature regime.
Here the behavior is controlled by the Ising energy and not by the microscopic time scale. 
\end{abstract}
\maketitle

The kinetic Ising model is a very simple but effective model to study non-equilibrium situations. 
The model is based on the Ising model which describes the interaction of a set of spins with 
values $\{\sigma _i = \pm 1 \}$ at each lattice site $i$. Starting from an arbitrary initial 
state, the simplest dynamics consist of a single spin-flip process $\sigma _i \rightarrow - \sigma _i$  
which is realized with a certain transition rate $w_i(\sigma _i)$. This rate is included into a Master 
equation, which is an equation of motion for the single-time probability $p(n,t)$. In our case the 
configuration $n$ consists of the set of all spins in $d$ dimensions. This is usually referred to as the Glauber model 
\cite{Glauber}. The problem is to find out an analytical expression for the transition rate. 
In order to ensure that the system eventually relaxes to an equilibrium state, one imposes the principle of detailed balance, 
and the transition rate is chosen in accordance with that principle. One choice is proposed in \cite{suku}
\begin{equation}
w_i^{G}(\sigma _i) = \frac{1}{2\alpha } \left[1 - \sigma _i \tanh\left(\frac{E_i }{T}\right)\right]\,.
\label{eq1}
\end{equation}
Here the temperature $T$ is given in terms of the Boltzmann constant and $E_i$ is the local energy of 
the Ising model 
\begin{equation}
E_i = h_i + \sum_{j(i)} J_{ij} \sigma _j\,,  
\label{eq.3}
\end{equation}
where $j(i)$ means the sum over all nearest neighbors of lattice site $i$. The local energy 
arises from the Hamiltonian given by  
\begin{equation}
H = -\sum_i h_i \sigma _i - \frac{1}{2} \sum_{i,j} J_{ij} \sigma _i \sigma _j\,.
\label{eq2}
\end{equation}
The quantity $h_i$ is an external field, and the summation in the interacting part goes 
over all pairs of nearest neighbor spins. Notice that the choice of the transition rate in 
Eq.~(\ref{eq1}) is not unique. 
Moreover, the interaction energy and the coupling to the heat bath with temperature $T$ is 
only incorporated into the Master equation via the principle of detailed balance. 
Let us stress that the different choices for the transition probability is discussed in the context 
of Monte Carlo methods \cite{bn}. 
The goal of the present note is to include the energy functional and the temperature 
directly into the Master equation from the beginning. To that aim we use a mapping of the Master 
equation onto a dynamic equation in terms of second quantized operators \cite{do,gr,pe,gwsp,satr}, 
for a recent review see \cite{gs}.\\
Let us start from a general Master equation written in the form 
\begin{equation}
\partial_t p(n,t) =\sum_{n'}\big[\,w(n\mid n')\, p(n',t) - w(n'\mid n)\, p(n,t)\,\big] 
\equiv \sum_{n'} L(n,n')\,p(n',t)\,.
\label{ma1}
\end{equation}
Here $p(n,t)$ is the probability that a certain configuration $n$ is realized at time $t$ and 
$w(n \mid n')$ plays the role of the transition probability per unit time from configuration $n$ to 
$n'$. In our case the configuration $n$ is given by the orientation of the set of spins. 
The principle of detailed balance means that the stationary distribution $p_s(n)$ fulfills 
\begin{equation}
\frac{p_s(n')}{p_s(n)} = \frac{w(n' \mid n)}{w(n \mid n')}\,.
\label{ma2}
\end{equation}
In case the static properties of the system are governed by a Hamiltonian let us make 
the following ansatz \begin{equation}
w(n'\mid n) =  \exp[-\beta H(n')/2]\,V(n'\mid n)\,\exp[\beta H(n)/2]\,.
\label{ma3}
\end{equation}
This ansatz is motivated by the conventional Arrhenius ansatz for transition rates, 
see also the result obtained in Eq.~(\ref{ev5}) and the comment made there.  
Further let us remark that we have assumed the validity of the commutator relation $[ H(n), H(n')] = 0 $. 
In Eq.~(\ref{ma3}) the parameter $\beta $ is an arbitrary one. In case the system is coupled 
to a heat bath we will identify $\beta $ with the inverse temperature $T$ in units of 
the Boltzmann constant. Then the stationary condition Eq.~(\ref{ma2}) can be rewritten as 
\begin{equation}
\frac{p_s(n')}{p_s(n)} = \frac{V(n'\mid n)}{V(n\mid n')}\exp(\beta [H(n)-H(n')])\,.
\label{ma4}
\end{equation}
Inserting Eq.~(\ref{ma3}) in the Master equation (\ref{ma1}) we get
\begin{eqnarray}
\partial_tp(n,t) &=& \sum_{n'} [\,\exp(-\beta H(n)/2)\,V(n\mid n')\,\exp(\beta H(n')/2)\,p(n',t) \nonumber\\
&-& \exp(-\beta H(n')/2) V(n'\mid n) \exp(\beta H(n)/2)\,p(n,t)\,]\,.
\label{ma5}
\end{eqnarray}
Now the further aim is to rewrite Eq.~(\ref{ma5}) using second quantized operators. Following 
Refs. \cite{do,gr,pe,satr,schutr1,schutr2}, the probability distribution $p(n,t)$ can 
be related to a state vector $\left| F(t)\right\rangle $ in a 
Fock-space according to $p(n,t)=\left\langle n|F(t)\right\rangle $ and 
$\left| F(t)\right\rangle =\sum_{n}p(n,t)\left| n\right\rangle $, respectively, where the basic vectors 
$\left| n \right\rangle $ can be expressed by second quantized operators. Using this representation, the 
underlying Master equation can be transformed into an equivalent evolution equation in a Fock-space, 
written in the form:  
\begin{equation}
\partial _{t}\left| F(t)\right\rangle = \hat{L} \left| F(t)\right\rangle\,.  
\label{ev1}
\end{equation}
The dynamical matrix elements $L(n,n^{\prime })$ within the Master equation are mapped onto the operator 
$\hat{L} = \hat{L}(a,a^{\dagger })$, where $a$ and $a^{\dagger }$ are the annihilation and creation operators, respectively. 
Here the matrix elements of the operator $\hat{L}(a,a^{\dagger })$ coincide with the matrix elements 
$L(n,n^{\prime })$.
Originally, this transformation had been applied for the Bose case with unrestricted occupation numbers 
\cite{do,gr,pe}. Here, we consider the case of restricted occupation numbers \cite{gwsp,satr,schutr1,schutr2}. 
In order to preserve the restriction of the occupation number in the underlying dynamical equations, 
the commutation rules of the operators $a$ and $a^{\dagger }$ are chosen as Pauli operators \cite{satr,gwsp,al}: 
\begin{equation}
\lbrack a_{i},a_{j}^{\dagger }]=\delta _{ij}(1-2a_{i}^{\dagger }a_{i})\,,\quad
\quad [a_{i},a_{j}]=[a_{i}^{\dagger },a_{i}^{\dagger }]=0\,,\quad \quad
a_{i}^{2}=(a_{i}^{\dagger })^{2}=0\,.
\label{ev2}
\end{equation}
The relation to the spin variable is $\sigma_i = 1 - 2 a_{i}^{\dagger}\,a_{i}$. In case of a single spin-flip 
process the evolution operator $\hat{L}$ reads
\begin{equation}
\hat{L} = \sum_i \left[ \lambda\, (1 - a_i^{\dagger} )\,a_i + \gamma\, (1 - a_i)\, a_i^{\dagger}\right]\,,
\label{ev3c}
\end{equation}
where $\lambda$ and $\gamma$ are the temperature dependent transition rates, the determination 
of those is beyond the scope of the present approach. The flip-rates are assumed in accordance to the 
principle of detailed balance manifested in Eq.~(\ref{ma2}). Obviously, the transition rates should 
depend on the details of the mutual interaction of the spins. 
Therefore the evolution operator should be extended by including the temperature and the interaction. In 
accordance with Eq.~(\ref{ma5}) we propose the following generalization 
\begin{equation}
\hat{L} = \kappa \sum_i \left[(1 - a_i^{\dagger} ) \exp(-\beta H /2) a_i \exp(\beta H/2 ) +  
(1 - a_i) \exp(-\beta H / 2 ) a_i^{\dagger} \exp(\beta H /2 )\right]\,.
\label{ev3a}
\end{equation}
Here, $\kappa $ is a parameter which fixes the time scale of the flip-process and $H$ is the Hamiltonian 
for the underlying interaction given by Eq.~(\ref{eq2}). 
Because of the relation between $\sigma _i$ and the annihilation and creation operators   
the evolution operator can be rewritten in terms of these operators as 
\begin{eqnarray}
\hat{L} &=& \kappa \sum_i \left[(1 - a_i^{\dagger} )\, a_i \exp\left(\frac{E_i}{T}\right) +  
(1 - a_i)\, a_i^{\dagger} \exp\left(-\frac{ E_i}{T} \right)\right] \nonumber\\
\mbox{with}\quad E_i &=& h_i + J(0) - 2 \sum_{j(i)} J_{ij} a_j^{\dagger} a_j\,;\quad J(0) = \sum_i J_{ij}\,.
\label{ev5}
\end{eqnarray}
Notice that the last relation is derived only by using the algebraic properties of the operators. 
In case of vanishing mutual interaction, i.e. $J = 0$, one observes that the last relation is equivalent to 
the conventional Arrhenius ansatz. To that aim the comparison of Eqs.~(\ref{ev3c}) and (\ref{ev5}) 
yields to the identification $\lambda = \kappa \exp(\,h/T\,)$ and $\gamma = \kappa \exp(-h/T\,)$ in according to the Arrhenius ansatz. Our 
quantum approach yields therefore a the possibility to formulate the Arrhenius ansatz in a more 
formal and mathematical manner.\\ 
To proceed further we follow Doi \cite{do} and calculate the average of an arbitrary physical quantity 
$B(n) $ by using the average of the corresponding operator 
$B = \sum_{n}\left| n\right\rangle B(n)\left\langle n\right| $ 
via \cite{schutr0} 
\begin{equation}
\left\langle B(t)\right\rangle = \sum_{n} p(n,t)B(n)= 
\left\langle s\left| B \right| F(t)\right\rangle.  
\label{ev2a}
\end{equation}
Here we have used the projection state 
$\left\langle s\right| =\sum_{n}\left\langle n\right| $, which is realized only for  
spin $1/2$ fermions in such a simple form. The normalization condition for the 
probability density is included in the condition $\left\langle s|F(t)\right\rangle =1$ with the consequence 
\cite{schutr0}, that the evolution operator fulfills always the relation $\langle s|\,L = 0$.
In the present case the averaged spin-variable obeys 
\begin{equation}
\frac{1}{2\kappa } \frac{\partial}{\partial t} \langle \sigma _i \rangle = 
\left\langle \sinh\left(\frac{E_i}{T}\right) \right\rangle 
- \left\langle \sigma _i \cosh\left(\frac{E_i}{T}\right)\right\rangle \,.
\label{ev3b}
\end{equation}  
Based upon the conventional Master equation Eq.~(\ref{ma1}) the averaged spin satisfies 
\begin{equation}
\frac{\partial}{\partial t} \langle \sigma _i \rangle = - 2 \langle\, \sigma _i\, w_i(\sigma _i)\, \rangle\,,
\label{ev4}
\end{equation}
where due to Glauber \cite{Glauber} or Suzuki et al. \cite{suku} the single transition rate  
$w_i(\sigma _i) \equiv w(\sigma _1, \dots, \sigma _i\,, \dots, \sigma _N \mid \sigma _1, \dots, -\sigma _i\,, 
\dots, \sigma _N)$  
is heuristically chosen in Eq.~(\ref{eq1}). This form of the transition rate is not 
uniquely determined by the principle of detailed balance. In our approach we can directly find the 
transition rate by applying the relation
$$
\sigma _i \exp\left(-\frac{E_i \sigma _i}{T}\right) = \sigma _i \cosh\left(\frac{E_i}{T}\right) - 
\sinh\left(\frac{E_i}{T}\right)\,.
$$
Using this relation and Eqs.~(\ref{ev3b},\ref{ev4}) we get immediately
\begin{equation}
w_i (\sigma _i) = \kappa \exp(-\frac{E_i \sigma _i}{T})\equiv \kappa \cosh\left(\frac{E_i }{T}\right)
\left[1 - \sigma _i \tanh\left(\frac{E_i}{T}\right)\right]\,.
\label{ev6}
\end{equation}
The transition rate is related to the heuristic one by assuming that the time scale 
$\alpha $ in Eq.~(\ref{eq1}) is controlled by the spin configuration as well as the temperature. 
Now let us discuss the transition rate obtained by Eq.~(\ref{ev6}) in detail, especially for the 
case of a vanishing external field. In that case the local energy is reduced to 
$E_i =  \sum_{j(i)}J_{ij} \sigma _j\,,$
where the summation is performed over all the $z$ nearest neighbors of the lattice site $i$. In the high 
temperature limit both transition probabilities $w_i^{G}$ and $w_i$ does not distinguish. In that case the 
transition rate is obviously only determined by the microscopic time scale $\alpha $ or $\kappa ^{-1}$. 
Both probabilities are independent on the spin configuration. In the low temperature limit $T \ll E_i$ the 
situation is completely different. Let us firstly assume that the spin at site $i$ is directed upwards, i.e. 
$\sigma _i = 1$. 
Then one has to distinguish two cases:

(i) $E_i < 0$: In this realization the majority of nearest neighbor spins is not adapted to the preferred 
upward direction of the spin $\sigma _i$. With other words, the local spin configuration around $\sigma _i$ is unfavorable. 
As a consequence the transition probabilities behave like 
\begin{eqnarray}
\lim_{T \to 0} w_i\,(\sigma _i = 1) & = & \kappa \exp (\mid E_i \mid/T) \nonumber\\   
\lim_{T \to 0} w_i^{G}\,(\sigma _i = 1) &= & \frac{1}{\alpha }\,. 
\label{ev7}
\end{eqnarray}
While in the Glauber transition rate the microscopic time scale $\alpha ^{-1}$ plays the role of a lower cut-off, 
i.e. spin-flips are possible within this time scale, in our realization the transition rate is controlled by the 
by the energy $E_i$ of the Ising model. The rate increases drastically and leads to an immediate flip-process 
of the "wrong" spins. The time scale $\tau _i$ for a flip is of the order 
$$
\tau _i \approx \kappa ^{-1} \exp (- \mid E_i \mid/T)\,.
$$
The spins in the energetically unfavorable direction perform the flips with a very high rate in a quite short 
time interval.

(ii) $E_i > 0$: In that case the majority of the spins around lattice site $i$ are already adapted and the 
transition rate tends to zero according to 
\begin{equation} 
w_i (\sigma _i = 1) = \kappa \exp (-E_i/T)\,.
\label{ev8}
\end{equation}
A similar behavior is also observed within the approximation due to Glauber.\\
In the second case we consider the downward orientation $\sigma _i = -1$. As above we study the two cases of 
positive and negative local energy $E_i$. For $E_i < 0$ we get in the low temperature regime
\begin{equation}
\lim_{T \to 0} w_i\,(\sigma _i = - 1) = \kappa  \exp (-\mid E_i \mid/T) \to 0\,, 
\label{ev9}
\end{equation} 
The transition rate tends to zero, i.e. 
all spin flips are suppressed completely. In the opposite case $E_i > 0$ it results 
\begin{equation}
\lim_{T \to 0} w_i\,(\sigma _i = - 1) = \kappa  \exp ( E_i /T)\,,
\label{ev10}
\end{equation}
which supports a very high flip rate, whereas the Glauber rate remains simply constant.\\

\noindent Summarizing our brief report, we have revisited the well established kinetic Ising model 
in terms of second quantized operators. The coupling to a heat bath at temperature $T$ and the 
underlying interaction are included similar to the Heisenberg-picture of operators. As the result 
we get an exact expression for the transition rate which is likewise in accordance to 
detailed balance. The differences to the conventional Glauber model consists in the low temperature regime. 
Here the transition rate is controlled by the Ising energy and not by the microscopic time scale. 
For low temperatures the transition rate is better adapted to the physical situation in mind.    

\begin{acknowledgments}
This work has been supported by the DFG: SFB 418 and SFB 569.
\end{acknowledgments}

\end{document}